\documentclass{ifacconf}

\usepackage{graphicx}      % include this line if your document contains figures
\usepackage{natbib}        % required for bibliography
\usepackage{pgfplots}
\usepackage{filecontents}
\usepackage{environ}
\usepackage{epsfig} 
\usepackage{amsmath,nccmath} 
\usepackage{amssymb} 
\usepackage{blindtext}
\usepackage{tikz}
\usetikzlibrary{graphs,graphs.standard}
\usetikzlibrary{arrows,shapes,positioning}
\usetikzlibrary{decorations.markings,decorations.pathmorphing,
	decorations.pathreplacing}
\usetikzlibrary{calc,patterns,shapes.geometric}
\usetikzlibrary{arrows}
\usetikzlibrary[arrows]
\usetikzlibrary{shapes.misc}
\usetikzlibrary{calc}
\usepackage{algorithm}
\usepackage{algorithmic}
\usepackage{balance}
\usepackage{subcaption}
\usepackage[mathscr]{eucal}
\let\mathcal\mathscr

\newtheorem{theorem}{Theorem}
\newtheorem{assumption}{Assumption}
\newtheorem{corollary}{Corollary}
\newtheorem{remark}{Remark}
\newtheorem{example}{Example}
\newtheorem{proposition}{Proposition}

\def\QED{\hfill{\mbox{\rule[0pt]{1.3ex}{1.3ex}}}} % for a filled box

\begin{document}
\begin{frontmatter}

\title{Control~of~Networked~Systems~by~Clustering: The Degree of Freedom Concept\thanksref{footnoteinfo}} 

\thanks[footnoteinfo]{This project has received funding from the European Research Council (ERC) under the European Union’s Horizon 2020 research and innovation programme grant agreement OCAL, No. 787845.}

\author[First]{Andrea Martinelli}
\author[First]{John Lygeros} 

\address[First]{Automatic Control Laboratory (IfA), ETH Zurich \\ Physikstrasse 3, 8092 Zurich, Switzerland \\ {\{andremar,lygeros\}@control.ee.ethz.ch}}

\begin{abstract}                % Abstract of not more than 250 words.
We address the problem of local flux redistribution in networked systems. The aim is to detect a suitable cluster which is able to locally adsorb a disturbance by means of an appropriate redistribution of control load among its nodes, such that no external node is affected. Traditional clustering measures are not suitable for our purpose, since they do not explicitly take into account the structural conditions for disturbance containment. We propose a new measure based on the concept of \textit{degree of freedom} for a cluster, and we introduce a heuristic procedure to quickly select a set of nodes according to this measure. Finally, we show an application of the method in the context of DC microgrids voltage control.
\end{abstract}


\end{frontmatter}
%===============================================================================

\section{INTRODUCTION}\label{Section Introduction}

In many networked systems, the uncontrolled propagation of local disturbances may lead to catastrophic effects on the dynamics of the whole network (\cite{SandellDecentralizedControl,HespanhaNetworkedControl}). Propagation of disturbances can deteriorate performance or even invalidate the stability achieved by the local loops. For these reasons, it is often necessary to design a higher-layer architecture that guarantees stability and performance satisfaction on a global scale. Our idea is to develop such an architecture that would commission the disturbance absorption only to a suitable subset of nodes to locally limit the propagation. Since this requirement relates to the connections among nodes encoded in the graph of the network, one would expect graph clustering methods to form the basis for determining which collection of nodes to assign the task of locally containing the disturbances to. Standard graph clustering techniques are based, amongst others, on intra/inter cluster link density (e.g. modularity), escape probability of a random walker (e.g. persistence probability), or eigenvector analysis of the Laplacian matrix (spectral analysis), as reported in \cite{SchaefferGraphClustering} and \cite{FortunatoCommunityDetection}. Other approaches to disturbance suppression in dynamical networks involve the concepts of time-scale separation (\cite{ChowTimeScaleModeling}), controllability Gramian (\cite{ISHIZAKI2015238}) and optimal resource allocation (\cite{PreciadoOptimalControlAllocation}), to cite a few. None of these graph properties and methods, however, capture the conditions that we want to impose for disturbance containment. \\
In analogy with other disciplines such as classical mechanics and statistics we introduce a new quantity, the \textit{degree of freedom (dof)}, that measures the structural availability of the cluster to locally contain a disturbance. This measure is also linked with the ability to redistribute internally the control load without affecting external nodes. The evaluation of the \textit{dof} only requires to compute the rank of a certain submatrix of the Laplacian. The idea is to both develop global clustering algorithms, that search for the optimal graph partition according to our measure, and local greedy algorithms, that detect the best local cluster to contain the propagation. 

Voltage/current control in microgrids will be the application domain to test the functionality of our methods. To the best of the authors' knowledge, most of the literature in this area focuses on nominal stability and scalability properties of the control architectures (\cite{Dragicevic1,Meng,Michele}), often without explicitly considering network disturbances and control saturations. Existing secondary solutions aim at reaching current sharing through consensus-based techniques (\cite{CucuzzellaConsensus,MicheleConsensusAutomatica}), but (i) every node in the network is involved in the references adjustment and (ii) saturations and disturbances are not taken into account. The introduction of higher-level schemes to locally contain disturbances in such models can be regarded as a safe practice to guarantee voltage/current stability during microgrids operation. 

The contributions of this paper can be summarized as follows. First of all, we define a new measure to evaluate the quality of a cluster, which is based on the original concept of degree of freedom. To strengthen the definition of \textit{dof}, we prove that, under clustering assumptions, any leading principal matrix of a Laplacian is nonsingular. Then, we introduce a greedy algorithm to detect a cluster according to our measure. Finally, we show the benefit of this framework in the area of microgrids voltage control.

%After introducing useful background theory and notation in Section \ref{Section basics}, we present in Section \ref{Section DOF} the concept of degree of freedom for a cluster. Then, in Section \ref{Section Algorithm}, we introduce the network model and we show a greedy algorithm to detect clusters according to our measure. Section \ref{Section Application} is devoted to the application of the developed method to DC microgids voltage regulation. In the last Section we draw conclusions and discuss about potential extensions.

%%%%%%%%%%%%%%%%%%%%%%%%%%%%%%%%%%%%%%%%%%%%%%%%%%%%%%%%%%%%%%%%%%%%%%%%%%%%%%

\section{BASICS}\label{Section basics}

In this Section we introduce concepts from graph theory and matrix theory that will be used for subsequent proofs, and will help to understand the notation throughout the remainder of the paper. The interested reader may refer to the textbooks by \cite{GodsilGraphTheory} and \cite{HornJohnsonMatrix} and the references therein for an in-depth discussion of the following concepts on graph and matrix theory. In general, for the sake of consistency, we denote graphs, sets, matrices and vector spaces with bold ($\mathbf{G}$), calligraphic ($\mathcal{P}$), Roman ($L$) and blackboard bold ($\mathbb{R}$) letters, respectively.

\subsection{Graph Theory}

A simple undirected graph is a pair $\mathbf{G} = (\mathcal{V},\mathcal{E})$, where $\mathcal{V} = \{ 1,\ldots,n \}$ is the node set and $\mathcal{E} \subseteq \mathcal{V} \times \mathcal{V}$ is the edge set. For each node $i \in \mathcal{V}$, $\mathcal{N}_i$ denotes the set of its neighbors. The degree of a node $i$ is the number of its neighbors, and it is denoted $\mbox{deg}(i) = |\mathcal{N}_i|$. Graph topology can be expressed by means of the adjacency matrix $A = [a_{ij}] \in \mathbb{R}^{n \times n}$, where $a_{ij} = 1$ if $(i,j) \in \mathcal{E}$, and $a_{ij} = 0$ otherwise. A convenient alternative representation is given by the Laplacian matrix $L = D - A \in \mathbb{R}^{n \times n}$, where $D = [d_{ij}]$ is the diagonal matrix with $d_{ij} = \mbox{deg}(i)$ if $j=i$, and $d_{ij} = 0$ otherwise. An undirected graph is connected if there is a path between every pair of nodes. An induced subgraph is formed from a subset of the nodes of the graph, and all the edges connecting pairs of nodes in that subset.

\subsection{Matrix Theory}

Let $\mathcal{M}$ be the set of all real matrices, and $\mathcal{M}_n$ the set of all real $n$-by-$n$ matrices. 

\begin{defn}[Leading principal matrix]
	The leading principal matrix of order $p \le n$ of a matrix $A \in \mathcal{M}_n$ is the square upper-left submatrix of $A$ obtained by removing the last $n - p$ rows and columns from $A$.
\end{defn}

\begin{defn}[Permutation matrix]
	A matrix $P \in \mathcal{M}_n$ is a \textit{permutation matrix} if exactly
	one entry in each row and column is equal to 1 and all other entries are 0.
\end{defn}

\begin{defn}[Irreducibility]
	A matrix $A \in \mathcal{M}_n$ is \textit{reducible} if there exists a permutation matrix $P$ such that $P'AP$ is a block upper-triangular matrix. If it is not the case, matrix $A$ is \textit{irreducible}. \\
	It can be shown that if $A$ is the adjacency (or Laplacian) matrix of a graph $\mathbf{G}$, then $A$ is irreducible if and only if $\mathbf{G}$ is connected (Theorem 6.2.24 in \cite{HornJohnsonMatrix}).
\end{defn}

\begin{defn}[Diagonal dominance]
	A matrix $A = [a_{ij}]$ $\in \mathcal{M}_n$ is diagonally dominant if
	\[ |a_{ii}| \ge \sum_{j \ne i}|a_{ij}| \quad \mbox{for all} \quad i = 1,\ldots,n. \] 
\end{defn}
\cite{TausskyIrreducibleMatrix} strengthened the Levy-Desplanques theorem on nonsingularity of strictly diagonally dominant matrices to the irreducible case:

\begin{theorem}[Taussky]\label{Taussky}
	Let $A = [a_{ij}] \in \mathcal{M}_n$ be an (i) irreducible and (ii) diagonally dominant matrix. If (iii) $\exists \: i \in \{ 1,\ldots,n \}$ such that
	$ |a_{ii}| > \sum_{j \ne i}|a_{ij}|, $
	then $A$ is nonsingular. If, in addition, every diagonal entry of $A$ is positive, then every eigenvalue of $A$ has positive real part.
\end{theorem}

%%%%%%%%%%%%%%%%%%%%%%%%%%%%%%%%%%%%%%%%%%%%%%%%%%%%%%%%%%%%%%%%%%%%%%%%%%%%%%

%% There are a number of predefined theorem-like environments in
%% ifacconf.cls:
%%
%% \begin{thm} ... \end{thm}            % Theorem
%% \begin{lem} ... \end{lem}            % Lemma
%% \begin{claim} ... \end{claim}        % Claim
%% \begin{conj} ... \end{conj}          % Conjecture
%% \begin{cor} ... \end{cor}            % Corollary
%% \begin{fact} ... \end{fact}          % Fact
%% \begin{hypo} ... \end{hypo}          % Hypothesis
%% \begin{prop} ... \end{prop}          % Proposition
%% \begin{crit} ... \end{crit}          % Criterion

\section{GRAPH CLUSTERING BASED ON A DEGREE OF FREEDOM MEASURE}\label{Section DOF}

\begin{figure}
		\centering
		\begin{tikzpicture}[auto,
			node_style/.style={circle,draw,minimum width=4mm,fill=blue!15!},
			edge_style/.style={draw=black},]
			
			\node[node_style] (v1) at (-0.5,1) {};
			\node[node_style] (v2) at (-0.5,2) {};
			\node[node_style] (v3) at (1.5,2) {};
			\node[node_style] (v4) at (0.5,1) {};
			\node[node_style] (v5) at (0.5,0) {};
			\node[node_style] (v6) at (1.5,0.5) {};
			
			\node at (v1) {\scriptsize$1$};
			\node at (v2) {\scriptsize$2$};
			\node at (v3) {\scriptsize$3$};
			\node at (v4) {\scriptsize$4$};
			\node at (v5) {\scriptsize$5$};
			\node at (v6) {\scriptsize$6$};
			
			\draw[edge_style]  (v1) edge node{} (v2);
			\draw[edge_style]  (v2) edge node{} (v3);
			\draw[edge_style]  (v2) edge node{} (v4);
			\draw[edge_style]  (v3) edge node{} (v4);
			\draw[edge_style]  (v4) edge node{} (v6);
			\draw[edge_style]  (v4) edge node{} (v5);
			
			\draw[dashed] (-0.5,1.5) ellipse (0.4cm and 1cm);
			\draw[dashed,rotate around={-25:(1.1,0.9)}] (1.1,0.9) ellipse (0.9cm and 1.5cm);
			
			\node at (-1.4,1.75) {$\mathbf{G}:$};
			\node (C1) at (-0.7,-0.2) {$\mathcal{C}_1$};
			\node (C2) at (2.2,-0.2) {$\mathcal{C}_2$};
			\draw [very thin] (C1) to[in=270,out=90] (-0.5,0.5);
			\draw [very thin] (C2) to[in=240,out=110] (1.8,0.3);

			\node[node_style] (v12) [right=4cm of v1] {};
			\node[node_style] (v22) [right=4cm of v2] {};
			\node[node_style] (v32) [right=4.7cm of v3] {};
			\node[node_style] (v42) [right=4.7cm of v4] {};
			\node[node_style] (v52) [right=4.7cm of v5] {};
			\node[node_style] (v62) [right=4.7cm of v6] {};
			
			\node at (v12) {\scriptsize$1$};
			\node at (v22) {\scriptsize$2$};
			\node at (v32) {\scriptsize$3$};
			\node at (v42) {\scriptsize$4$};
			\node at (v52) {\scriptsize$5$};
			\node at (v62) {\scriptsize$6$};
			
			\draw[edge_style]  (v12) edge node{} (v22);
			\draw[edge_style]  (v32) edge node{} (v42);
			\draw[edge_style]  (v42) edge node{} (v62);
			\draw[edge_style]  (v42) edge node{} (v52);
			
			%	\draw[dashed] (-0.5,1) ellipse (0.75cm and 1.8cm);
			%	\draw[dashed,rotate around={-27:(2.25,0.5)}] (2.25,0.5) ellipse (1.4cm and 2.25cm);
			
			\node at (3.2,1.75) {$\mathbf{G}_1:$};
			\node at (5.1,1.75) {$\mathbf{G}_2:$};
			\end{tikzpicture}
	\caption{A partition $\mathcal{P} = \{ \mathcal{C}_1,\mathcal{C}_2 \}$ is displayed on the undirected graph $\mathbf{G}$. According to Assumption \ref{induced subgraph}, the induced subgraphs $\mathbf{G}_1$ and $\mathbf{G}_2$ are connected.}
	\label{Example graph}
\end{figure}
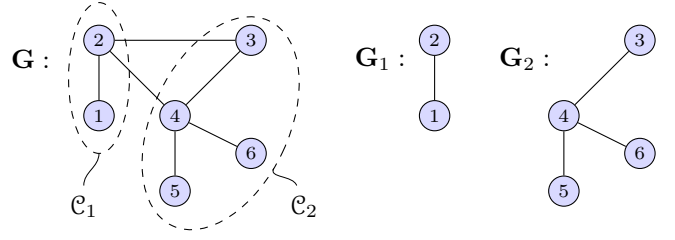

In this Section, we propose a measure to evaluate the quality of a cluster, or a partition, which is based on the concept of \textit{degree of freedom (dof)}. The aim is to partition an undirected graph $\mathbf{G} = (\mathcal{V},\mathcal{E})$ with $n$ nodes into $m$ non-overlapping clusters. Let the partition be denoted by $\mathcal{P} = \{ \mathcal{C}_1, \mathcal{C}_2, \ldots, \mathcal{C}_m \}$. Each cluster $\mathcal{C}_i$ contains a certain subset of nodes such that
\begin{equation*}
\bigcup\limits_{i=0}^{m} \mathcal{C}_{i} = \mathcal{V}, \quad \mbox{and} \quad \mathcal{C}_{j} \cap \mathcal{C}_k = \emptyset \quad \forall j \ne k.
\end{equation*}
We introduce two assumptions on the graph $\mathbf{G}$ and the partition $\mathcal{P}$.
\begin{assumption}\label{strong;y connected}
	The graph $\mathbf{G}$ is connected.
\end{assumption}
\begin{assumption}\label{induced subgraph}
	The induced subgraph identified by the nodes in each cluster $\mathcal{C}_i \in \mathcal{P}$, denoted by $\mathbf{G}_i$, is connected. Moreover, to avoid trivial clusters, we assume that $m \ge 2$ and $\mathcal{C}_i \neq \emptyset \;\: \forall i = 1,\ldots,m$.
\end{assumption}%
%Assumption \ref{induced subgraph} guarantees that each cluster is a connected graph, i.e. there are no isolated nodes. Moreover, we are specifying that the subgraph identified by $\mathcal{C}_i$ contains all the edges of the original graph $\mathbf{G}$ for which both endpoints belong to nodes in $\mathcal{C}_i$.
If we represent the topology of $\mathbf{G}$ with the corresponding Laplacian matrix $L$, the choice of a partition $\mathcal{P}$ induces a block partition of $L$:

\begin{equation}\label{laplacian block partition}
L = \left( \begin{tabular}{c|c|c|c}
$C_1$ & $Y_{12}$ & $\cdots$ & $Y_{1m}$  \\
\hline
$Y_{12}^{'}$ & $C_2$ &  &   \\
\hline
\vdots &  & $\ddots$ & \\
\hline
$Y_{1m}^{'}$ &  &  &  $C_m$
\end{tabular}\right).
\end{equation}

We will call the matrix $C_i$ the \textit{diagonal block} associated with the cluster $\mathcal{C}_i$, and matrix $Y_i = \{ Y_{i1} | \cdots | Y_{im} \}$ the \textit{bridge matrix} of $\mathcal{C}_i$. As will soon become apparent, matrix $C_i$ carries information on how nodes in $\mathcal{C}_i$ are internally connected, and matrix $Y_i$ shows how $\mathcal{C}_i$ is linked to the rest of the graph.

\begin{theorem}\label{Theorem nonsingularity}
	Let $\mathbf{G}$ be a graph and $\mathcal{P}$ a partition satisfying Assumptions \ref{strong;y connected} and \ref{induced subgraph}, respectively, and let $L \in \mathcal{M}_n$ be the Laplacian matrix of $\mathbf{G}$. Then, any diagonal block of $L$ associated with a cluster in $\mathcal{P}$ is nonsingular and, in particular, has all eigenvalues with positive real part.
\end{theorem}

\begin{pf}
	Let $C_i$ be a diagonal block of $L$ corresponding to the cluster $\mathcal{C}_i$ in $\mathcal{P}$. The diagonal entries of $C_i$ are the degree of the nodes in the original graph $\mathbf{G}$, whereas off-diagonal entries correspond to the edges whose endpoints belong to nodes in $C_i$. The edges connecting nodes in $C_i$ to nodes in the rest of the graph are captured by the bridge matrix $Y_i = \{ Y_{i1} | \cdots | Y_{im} \}$, according to the block partition in \eqref{laplacian block partition}. As stated by Assumption \ref{induced subgraph}, we can define a new graph, $\mathbf{G}_i$, as the connected induced subgraph of $\mathbf{G}$ identified by the nodes in $C_i$. Then, we can decompose matrix $C_i$ as follows:
	\begin{equation}\label{decomposition}
	C_i = L_{i} + D_i,
	\end{equation}
	
	where $L_{i}$ is the Laplacian matrix that describes $\mathbf{G}_i$, and $D_i$ is the positive diagonal matrix whose entries are the degree deficiency of nodes in $\mathbf{G}_i$ with respect to the same nodes in $\mathbf{G}$. We know that $L_i$, since it is the description of the connected graph $\mathbf{G}_i$, is \textit{irreducible}. This means that there exists no \textit{permutation matrix} $P \in \mathcal{M}_p$ such that $P'L_iP$ is a block upper-triangular matrix. As pointed out in chapter 0.9.5 of \cite{HornJohnsonMatrix}, if $D_i$ is diagonal and $P$ is a permutation matrix, then $P'D_iP$ is diagonal as well. Therefore, there exists no permutation matrix $\tilde{P} \in \mathcal{M}_p$ such that 
	\[ \tilde{P}'C_i\tilde{P} = \tilde{P}'L_i\tilde{P} + \tilde{P}'D_i\tilde{P} \]
	
	is in block upper-triangular form. We conclude that matrix $C_i$ is irreducible. Moreover, $C_i$ is \textit{diagonally dominant} and there is at least one row where the magnitude of the diagonal element is strictly greater then the sum of the magnitudes of all other elements in that row. This is because the cluster identified by $\mathcal{C}_i$ is connected to at least one external node by Assumption \ref{strong;y connected}.
	
	As the conditions (i)-(ii)-(iii) of Theorem \ref{Taussky} (Taussky's Theorem) are satisfied, we can conclude that matrix $C_i$ is nonsingular. Furthermore, since all diagonal elements of $C_i$ are positive, then all eigenvalues of $C_i$ have positive real part. \QED
\end{pf}

\begin{corollary}
	A leading principal matrix of $L$ has all eigenvalues with positive real part if the induced subgraph identified by the nodes in the associated cluster is connected.
\end{corollary}

\begin{pf}
	Consider the leading principal matrix of $L$ of order $p < n$ to be the first diagonal block $C_1$, according to the partition \eqref{laplacian block partition}. Then, the results of Theorem \ref{Theorem nonsingularity} hold. \QED
\end{pf}

%\begin{remark}
%	Notice that Theorem \ref{Theorem nonsingularity} does not hold for the trivial cluster $m = 1$, i.e. when the diagonal block $C_i$ coincides with $L$ itself. In this case, condition (iii) of Taussky's Theorem does not hold anymore. In fact, by applying decomposition \eqref{decomposition}, we find $D_i = 0$, because the induced subgraph described by the nodes in $C_i$ coincides with $\mathbf{G}$ itself.
%\end{remark}

%\begin{rem}
%	Notice that Theorem \ref{Theorem nonsingularity} holds for each of the diagonal block matrices induced by the partition $\mathcal{P}$. In fact, according to block division in \eqref{laplacian block partition}, we can arbitrarily number the nodes of a cluster $\mathcal{C}_i$ to be the first $p$ nodes in the Laplacian, such that the corresponding matrix $C_i$ is a leading principal matrix of $L$ of order $p$.
%\end{rem}

\begin{defn}[Rank difference function] 
	Let us consider the function $\delta \, : \, \mathcal{M} \times \mathcal{M} \rightarrow \mathbb{N}$ that, given two matrices $M_1, M_2 \in \mathcal{M}$, returns the corresponding rank difference
	
	\begin{equation}
	\delta(M_1,M_2) = \mbox{rank}(M_1) - \mbox{rank}(M_2).
	\end{equation}
\end{defn}

We call $\delta$ the \textit{rank difference function}.

\begin{defn}[Degree of freedom of a cluster]
	Consider a graph $\mathbf{G}$ and a cluster $\mathcal{C}_i$ satisfying Assumptions \ref{strong;y connected} and \ref{induced subgraph}, respectively. We say that the cluster $\mathcal{C}_i$ has $d \in \mathbb{N}$ \textit{degrees of freedom} if $\delta(C_i,Y_i) = d$. That is, when the rank difference between the diagonal block $C_i$ and the corresponding bridge matrix $Y_i$ is equal to $d$. Moreover, we refer to the quantity $|\mathcal{C}_i| - \delta(C_i,Y_i)$ as the \textit{dof deficiency} of the cluster $C_i$.
\end{defn}

\begin{proposition}
	The function $\delta$, when applied to $C_i$ and $Y_i$, is restricted to the following interval
	
	\begin{equation}
		0 \le \delta(C_i,Y_i) < |\mathcal{C}_i| \quad \forall i = 1,\ldots m,
	\end{equation}
	
	that is, the \textit{dof} of $C_i$ are confined between zero and the cardinality of the cluster itself.
\end{proposition}

\begin{pf}
	Firstly, thanks to Theorem \ref{Theorem nonsingularity}, we know that matrix $C_i$ is nonsingular, and therefore full rank. On the other hand, since $C_i$ and $Y_i$ share the same number of rows, $\mbox{rank}(Y_i)$ cannot exceed the cardinality of $\mathcal{C}_i$. Formally,
	
	\begin{gather}
	\mbox{rank}(C_i) = |\mathcal{C}_i| \quad \land \quad \mbox{rank}(Y_i) \le |\mathcal{C}_i| \notag \\
	\implies \quad 0 \le \delta(C_i,Y_i) < |\mathcal{C}_i|. \label{dof limits}
	\end{gather}
	
	The strict inequality on the right hand side follows from the fact that, by Assumption \ref{strong;y connected}, $\mathbf{G}$ is connected. Hence, $\mbox{rank}(Y_i) > 0$, that concludes the proof. \QED
\end{pf}

\begin{remark}\label{Remark dof deficiency}
	We can provide an interpretation of \textit{dof} deficiency for a cluster, by stating that it represents the number of independent connections that exist between $\mathcal{C}_i$ and the rest of the graph. Since it is a rank-based measure, it does not simply count the total number of connections but, instead, only the links that connect $\mathcal{C}_i$ to the rest of the cluster in a ``different" way. 
\end{remark}
\begin{example}\label{example}
	
	Consider the graph $\mathbf{G}$, comprising $6$ nodes depicted in Fig. \ref{Example graph}. The graph is partitioned into two clusters,
	\begin{equation*}
	\mathcal{P} = \{ \mathcal{C}_1, \mathcal{C}_2 \} = \{ \{ 1,2 \}, \{ 3,4,5,6 \} \}.
	\end{equation*}
	
    Now we can represent the topology of $\mathbf{G}$ by means of the Laplacian matrix, emphasizing the block division induced by $\mathcal{P}$:
	\begin{align*}\label{laplacian example}
	L & = \left( \begin{array}{c|c}
	C_1 & Y_{12} \\
	\hline
	Y_{12}^{'} & C_2
	\end{array}\right) =\left( \begin{array}{cc|cccc}
	1 & -1 & 0 & 0 & 0 & 0 \\
	-1 & 3 & -1 & -1 & 0 & 0 \\
	\hline
	0 & -1 & 2 & -1 & 0 & 0 \\
	0 & -1 & -1 & 4 & -1 & -1 \\
	0 & 0 & 0 & -1 & 1 & 0 \\
	0 & 0 & 0 & -1 & 0 & 1 
	\end{array}\right).
	\end{align*}
	
The leading diagonal block $C_1$ can be further decomposed, according to \eqref{decomposition}, into 
	\begin{align*}
	C_1 = L_1 + D_1 = \left( \begin{array}{cc}
		1 & -1 \\ -1 & 1
		\end{array} \right) + \left( \begin{array}{cc}
		0 & 0 \\ 0 & 2
		\end{array} \right),
	\end{align*}
	and similarly for $C_2$. Matrix $L_1$ is the Laplacian that describes the induced subgraph $\mathbf{G}_1$ identified by the nodes in cluster $\mathcal{C}_1$, while $D_1$ is the diagonal matrix that contains the degree deficiency for each node in $\mathbf{G}_1$ with respect to $\mathbf{G}$. The two induced subgraphs are represented in Fig. \ref{Example graph}. Let us compute the \textit{dof} for the two clusters:
	
	\begin{align*}
	\delta(C_1,Y_{12})  & = \mbox{rank}(C_1) - \mbox{rank}(Y_{12}) = 2 - 1 = 1, \\
	\delta(C_2,Y_{12}^{'}) & = \mbox{rank}(C_2) - \mbox{rank}(Y_{12}^{'}) = 4 - 1 = 3.
	\end{align*}
	
	Notice that the degree deficiency of node 2 corresponds to the edges $(2,3)$ and $(2,4)$ of $\mathbf{G}$. We want to stress that the \textit{dof} computation for a cluster only requires the evaluation of one rank, the one associated to the bridge matrix $Y$. That is because, thanks to Theorem \ref{Theorem nonsingularity}, we know that the rank of the diagonal block corresponds to the cardinality of the cluster. 
	We can now get some intuition behind the meaning of \textit{dof}. Cluster $\mathcal{C}_1$ has 1 \textit{dof}. This is because we can modify the value of node 1 and redefine the flux exchange with 2, without altering how the cluster is perceived from the outside. Node 2, on the other hand, cannot be modified because it will have an impact on external nodes. This intuition will be more clear after the introduction of the coupling model \eqref{Coupling} in the next Section. Cluster $\mathcal{C}_2$ has a \textit{dof} deficiency of 1, even though both nodes 3 and 4 are involved in external connections. This is because they share a dependent connection, so it is possible to modify their values to redistribute flux within $\mathcal{C}_2$ as long as the net power injected into node 2 is zero. As it is pointed out by Remark \ref{Remark dof deficiency}, the \textit{dof} deficiency for a cluster counts the number of independent external connections.
\end{example}

%%%%%%%%%%%%%%%%%%%%%%%%%%%%%%%%%%%%%%%%%%%%%%%%%%%%%%%%%%%%%%%%%%%%%%%%%%%%%%

\section{CLUSTERING METHOD}\label{Section Algorithm}

In this Section, we firstly argue why standard clustering methods are not suitable to our purposes. Then, we introduce a network dynamical model and discuss about typical hierarchical control architectures. Finally, a greedy clustering algorithm based on the \textit{dof} concept is presented.

\subsection{Why are existing clustering methods not suitable?}

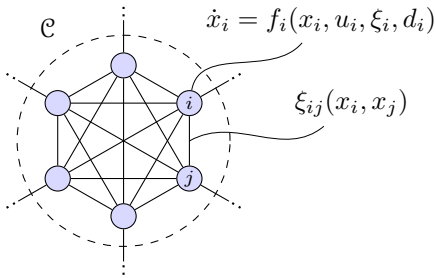
\begin{figure}
	\centering
	\scalebox{1}{
	\begin{tikzpicture}[auto,
	node_style/.style={circle,draw,fill=blue!15!},
	edge_style/.style={draw=black},]
	
	\graph { subgraph K_n [n=6,clockwise,radius=1cm,empty nodes,nodes={node_style}] };
	\draw[dashed] (0,0) ellipse (1.4cm and 1.4cm);
	
	\draw (1) to ($1.6*(1)$); \draw[dotted,thick] ($1.65*(1)$) to[] ($1.8*(1)$);
	\draw (2) to ($1.6*(2)$); \draw[dotted,thick] ($1.65*(2)$) to[] ($1.8*(2)$);
	\draw (3) to ($1.6*(3)$); \draw[dotted,thick] ($1.65*(3)$) to[] ($1.8*(3)$);
	\draw (4) to ($1.6*(4)$); \draw[dotted,thick] ($1.65*(4)$) to[] ($1.8*(4)$);
	\draw (5) to ($1.6*(5)$); \draw[dotted,thick] ($1.65*(5)$) to[] ($1.8*(5)$);
	\draw (6) to ($1.6*(6)$); \draw[dotted,thick] ($1.65*(6)$) to[] ($1.8*(6)$);
	
	\node at (2) {\scriptsize$i$};
	\node at (3) {\scriptsize$j$};
	\node at (-1,1.5) {$\mathcal{C}$};
	\node (eq) at (2.6,1.6) {$\dot{x}_i = f_i(x_i,u_i,\xi_i,d_i)$};
	\node (link) at (3,0.5) {$\xi_{ij}(x_i,x_j)$};
	
	\draw [very thin] (2) to[in=230,out=80] (eq);
	\draw [very thin] (0.87,0) to[in=220,out=30] (link);
	\end{tikzpicture}}
	\caption{The cluster $\mathcal{C}$ has zero \textit{dof}, even though it shows high internal and low external connection density.}
	\label{Complete graph}
\end{figure}

According to \cite{SchaefferGraphClustering}, measures for identifying clusters can be divided in (i) \textit{vertex similarity measures}, which assign values or properties to the nodes and then group them into consistent clusters, and (ii) \textit{fitness measures}, which define a function over the set of possible clusters and then choose those that optimize the function. Popular methods are based on intra/inter cluster connection density (modularity), escape probability of a random walker (persistence probability), or Laplacian eigenvectors (spectral analysis). To the best of our knowledge, none of these methods take into account the structural condition we require for the flux redistribution problem. 

Consider, for example, cluster $\mathcal{C}$ depicted in Fig. \ref{Complete graph}. Since the induced subgraph of $\mathcal{C}$ is a complete graph, it shows high internal and low external connection density. Moreover, a random walker has an escape probability of $1/6$ and a probability of remaining within the cluster of $5/6$. On the contrary, if we apply the \textit{dof} function $\delta$ to the cluster we can see that it has zero \textit{dof}, as any node modification will result in a variation of inter-cluster flux. 

\subsection{Flux redistribution problem}

In many applications, networked control problems are solved through hierarchical controllers that comprise multiple control layers, often operating at different timescales and with different control objectives (\cite{Vasquez,AlessioHierarchicalMPC}). For the sake of simplicity, we consider the case of a two-layer hierarchy. The bottom layer, called the primary layer, has the objective of tracking the reference values defined by the upper layer, called the secondary layer. The secondary layer is in charge of selecting the reference values to optimize some global performance measures. As summarised in Fig. \ref{Complete graph}, each node $i \in \mathcal{V}$ of the graph $\mathbf{G} = (\mathcal{V},\mathcal{E})$ hosts a dynamical system of the form
\begin{equation}\label{dynamical system}
\dot{x}_i = f_i\left( x_i,u_i,\xi_i,d_i \right), 
\end{equation} 

where $x_i \in \mathbb{R}^{N_i}$, $u_i \in \mathbb{R}$, $\xi_i \in \mathbb{R}$ and $d_i \in \mathbb{R}$ represent the vector of state variables, the input, the coupling and the disturbance of the dynamical system \eqref{dynamical system}. The interaction among nodes is described by a static function $\xi_{jk} = \xi_{jk}(x_j,x_k)$ on each link $(i,j) \in \mathcal{E}$, and the net flux injected into node $i$ is 
\begin{equation}\label{Coupling}
\xi_i = \sum_{j \in \mathcal{N}_i}\xi_{ij}.
\end{equation}  

Notice that we assume $u_i$, $\xi_i$ and $d_i$ to be scalar quantities and, to simplify the notation, we omit the time dependence of all variables in equations \eqref{dynamical system}-\eqref{Coupling}. The functions $f_i$ and $\xi_{jk}$ define the dynamics of \eqref{dynamical system} and the coupling between two neighbouring nodes, respectively. 

We assume that each dynamical system \eqref{dynamical system} is equipped with a local primary control law $u_i = g_i(x_i,x_i^{r},\xi_i,d_i)$ that, in nominal conditions, stabilizes the system around its reference $x_{i}^{r}$ defined by the secondary layer. Examples of such state-feedback stabilizing laws can be found, for instance, in \cite{Michele} and \cite{Cucuzzella}. When the references $x_i^{r}$ and disturbances $d_i$ are constant, the equilibrium condition can be expressed as 

\begin{equation}
	f_i\left( x_i,u_i,\xi_i,d_i \right) = \tilde{f}_i\left( x_i^{r},x_{j\in\mathcal{N}_i}^{r},d_i \right) = 0,
\end{equation}

where $\tilde{f}_i$ is used to stress that the set of references and disturbances are the only variables that define the equilibrium of the network. Note that the flux exchange between neighbors is exclusively determined by the choice of the references, i.e. $\xi_{ij} = \xi_{ij}(x_i^{r},x_j^{r})$. Any variation of the disturbances is locally compensated by the input $u_i$ through the map $g_i$, and a perfect track $x_i = x_i^{r}$ is achieved thanks to the stabilizing property of the primary loops.

Secondary control schemes are employed to manage the references $x_i^{r}$ in order to balance the control effort throughout the network, by means of consensus algorithms or reference value adjustments (\cite{MicheleConsensusAutomatica,Vasquez}). In our view, the major drawbacks of this approaches are: (i) the input saturations are usually not considered, (ii) other performance measures, such as power losses, are not considered, and (iii) every single node in the network is involved in the reference adjustment. In \cite{AndreaSecondaryControl}, a reference adjustment scheme is introduced where, at fixed time instants, the following centralized optimization problem is solved

\begin{equation}\label{optimization problem}
\begin{aligned}
\quad & & \underset{x^{r}}{\text{min}} \quad & V(x^{r}) \\
& &\mbox{s.t.} \quad & \tilde{f}_i\left( x_i^{r},x_{j\in\mathcal{N}_i}^{r},d_i \right) = 0 \quad \forall i   \\
& & & \textstyle \xi_i = \sum_{j \in \mathcal{N}_i}\xi_{ij} \quad \forall i   \\
& & & u_i \in [u_i^{\mbox{\footnotesize min}},u_i^{\mbox{\footnotesize max}}] \quad \forall i.
\end{aligned}	
\end{equation}

This method permits one to explicitly include input saturations and to define other performance measures (e.g. Joule heating, control balance,...) to be optimized with a global cost function $V$. \cite{AndreaSecondaryControl} consider a disturbance variation in a node $i$, referred to as the overloading node, identify a cluster which surrounds the overloading node by a simple algorithm, and solve a local version of the optimization problem \eqref{optimization problem} among the nodes in the cluster. The node exploration strategy used in that reference is based on the concept of \textit{k}-steps reachability set, where at each exploration step the cluster is enlarged with the nodes that are reachable in \textit{k} steps from the overloading node. The idea is simple to implement, but the major drawback is that a significant amount of nodes is usually involved in the cluster. The reason lies in the fact that the topology of the network is not exploited in the exploration process. In a sense, the objective of solving the flux redistribution as locally  as possible is penalized in favour of simplicity of execution. In the following, we introduce a clustering algorithm based on the \textit{dof} concept.

\subsection{Greedy clustering algorithm}

Algorithm 1 presents an exploration strategy which exploits graph topology, by searching those nodes that increase the \textit{dof} in the cluster. The cluster is initialized with the index of the overloading node $\mathcal{C} = \{ i \}$, and the neighbors of the cluster are $\mathcal{N}_{\mathcal{C}} = \mathcal{N}_i$. The initial \textit{dof} of the cluster is $\delta_{\mathcal{C}} = 0$, because any isolated node has zero \textit{dof}. At each time step, only one node is added to $\mathcal{C}$. Among the nodes in $\mathcal{H}$, namely the set of all neighbors of $\mathcal{C}$ that would increase the number of \textit{dof}, it is selected the one that maximises a certain \textit{availability measure} $\Psi_j$ associated to it (see equation \eqref{cluster update 1}). This measure can be designed so that it defines the capacity of $j$ to modify its reference value and control input without incurring saturations. An example of $\Psi$ is given in the next section. If there is no neighbor that would increase \textit{dof}, the algorithm selects the neighbor with the highest number of connections, to increase the probability that $\mathcal{H} \ne \emptyset$ at the next iteration. When a new node is added to $\mathcal{C}$, its set of neighbors is updated according to \eqref{neighbors update}, and a local version of the optimization problem \eqref{optimization problem} is solved. The procedure is iterated until a feasible solution $x^{r}_{\mathcal{C}}$ is found.

\begin{algorithm}
	\caption{\textit{Dof}-based clustering heuristic}
	\label{Algorithm}
	\begin{algorithmic}
		\STATE \textbf{given} overloading node $i$ with set of neighbors $\mathcal{N}_i $
		\STATE \textbf{initialize} $\mathcal{C} = \{ i \}$, $\mathcal{N}_{\mathcal{C}} = \mathcal{N}_i$, $\delta_{\mathcal{C}} = 0$
		\REPEAT \STATE $\mathcal{H} = \{ j \in \mathcal{N}_{\mathcal{C}} \; : \; \delta_{\mathcal{C},j} > \delta_{\mathcal{C}} \}$
		\IF {$\mathcal{H} \ne \emptyset$}
		\STATE 
		\setlength\abovedisplayskip{-10pt} \setlength\belowdisplayskip{1pt}
		\begin{flalign}\label{cluster update 1}
		& \textstyle \mathcal{C} \leftarrow \mathcal{C} \cup \big\{ j = \arg\max_{j \in \mathcal{H}} (\Psi_{j}) \big\} & 
		\end{flalign}
		\ELSE \STATE
		\setlength\abovedisplayskip{-10pt} \setlength\belowdisplayskip{1pt} 
		\begin{flalign}\label{cluster update 2}
		& \textstyle \mathcal{C} \leftarrow \mathcal{C} \cup \big\{ j = \arg\max_{j \in \mathcal{N}_{\mathcal{C}}} (\mbox{deg}(j)) \big\} &
		\end{flalign}
		\ENDIF
		\STATE 
		\setlength\abovedisplayskip{-10pt} \setlength\belowdisplayskip{1pt} \begin{flalign}\label{neighbors update}
		& \textstyle \mathcal{N}_{\mathcal{C}} \leftarrow (\mathcal{N}_{\mathcal{C}} \cup \mathcal{N}_j) \smallsetminus (j \cup (\mathcal{C} \cap \mathcal{N}_{j}))&
		\end{flalign}
		\STATE solve optimization problem \eqref{optimization problem} for nodes in $\mathcal{C}$
		\UNTIL a feasible solution to \eqref{optimization problem} is found
		\RETURN $x^{r}_{\mathcal{C}}$
	\end{algorithmic}
\end{algorithm}

%%%%%%%%%%%%%%%%%%%%%%%%%%%%%%%%%%%%%%%%%%%%%%%%%%%%%%%%%%%%%%%%%%%%%%%%%%%%%%

\section{APPLICATION TO DC MICROGRIDS}\label{Section Application}

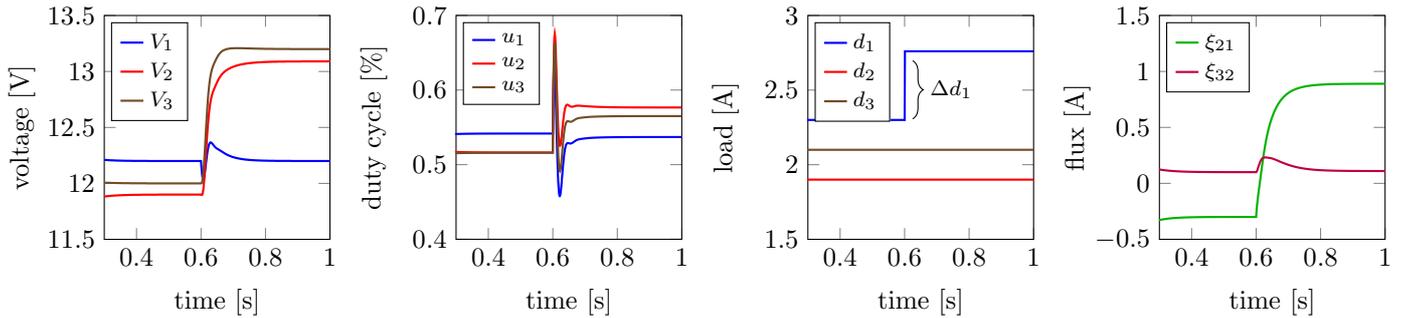
\begin{figure*}
	\hspace*{-0.18cm}    
	\begin{tikzpicture}
	\pgfplotsset{%
		width=0.25\linewidth,
		height=0.25\linewidth,
		legend image code/.code={
			\draw[mark repeat=2,mark phase=2]
			plot coordinates {
				(0cm,0cm)
				(0.15cm,0cm)        %% default is (0.3cm,0cm)
				(0.3cm,0cm)         %% default is (0.6cm,0cm)
			};%
		}
	}
	\begin{axis}[
	%only marks,                    % no lines
	every axis plot/.append style={thick},
	xmin=0.3, xmax=1,              % x-axis limits
	ymin=11.5, ymax=13.5,              % y-axis limits
	ylabel style={at={(axis description cs:0.05,0.5)}},
	xlabel={time [s]},      % x-axis label
	ylabel={voltage [V]},            % y-axis label
	legend style={font=\small},
	legend pos=north west,         % legend position on plot
	legend cell align=left,        % text alignment within legend
	%domain=20:180,                 % domain for plotted functions (not needed for scatter data)
	%samples=200,                   % plot 200 samples
	]
	\addplot+ [mark=none] table [col sep=comma,row sep=newline] {Simulation_V1.dat}; % add the first plot
	\addlegendentry{$V_1$}; % add the first plot's legend entry
	\addplot+ [mark=none] table [col sep=comma,row sep=newline] {Simulation_V2.dat};
	\addlegendentry{$V_2$};
	\addplot+ [mark=none] table [col sep=comma,row sep=newline] {Simulation_V3.dat};
	\addlegendentry{$V_3$};
	%\addlegendentry{Sammon Mapping};
	%\addplot[mark=triangle,green] {x^2/200 + rand*x/1.5};
	%\addlegendentry{Squared Stress};
	\end{axis}
	%\node at (1,2.4) {\small$V_1$};
	%\node at (1.6,2) {\small$V_2$};
	%\node at (1,2.4) {\small$V_3$};
	\end{tikzpicture}%
	~%
	\begin{tikzpicture}
	\pgfplotsset{%
		width=0.25\linewidth,
		height=0.25\linewidth,
		legend image code/.code={
			\draw[mark repeat=2,mark phase=2]
			plot coordinates {
				(0cm,0cm)
				(0.15cm,0cm)        %% default is (0.3cm,0cm)
				(0.3cm,0cm)         %% default is (0.6cm,0cm)
			};%
		}
	}
	\begin{axis}[
	%only marks,                    % no lines
	every axis plot/.append style={thick},
	xmin=0.3, xmax=1,              % x-axis limits
	ymin=0.4, ymax=0.7,              % y-axis limits
	ylabel style={at={(axis description cs:0.05,0.5)}},
	xlabel={time [s]},      % x-axis label
	ylabel={duty cycle [\%]},            % y-axis label
	legend style={font=\small},
	legend pos=north west,         % legend position on plot
	legend cell align=left,        % text alignment within legend
	%domain=20:180,                 % domain for plotted functions (not needed for scatter data)
	%samples=200,                   % plot 200 samples
	]
	\addplot+ [mark=none] table [col sep=comma,row sep=newline] {Simulation_D1.dat}; % add the first plot
	\addlegendentry{$u_1$}; % add the first plot's legend entry
	\addplot+ [mark=none] table [col sep=comma,row sep=newline] {Simulation_D2.dat};
	\addlegendentry{$u_2$};
	\addplot+ [mark=none] table [col sep=comma,row sep=newline] {Simulation_D3.dat};
	\addlegendentry{$u_3$};
	%\addlegendentry{Sammon Mapping};
	%\addplot[mark=triangle,green] {x^2/200 + rand*x/1.5};
	%\addlegendentry{Squared Stress};
	\end{axis}
	\end{tikzpicture}%
	~%
	\begin{tikzpicture}
	\pgfplotsset{%
		width=0.25\linewidth,
		height=0.25\linewidth,
		legend image code/.code={
			\draw[mark repeat=2,mark phase=2]
			plot coordinates {
				(0cm,0cm)
				(0.15cm,0cm)        %% default is (0.3cm,0cm)
				(0.3cm,0cm)         %% default is (0.6cm,0cm)
			};%
		}
	}
	\begin{axis}[
	%only marks,                    % no lines
	every axis plot/.append style={thick}, 
	xmin=0.3, xmax=1,              % x-axis limits
	ymin=1.5, ymax=3,              % y-axis limits
	ylabel style={at={(axis description cs:0.05,0.5)}},
	xlabel={time [s]},      % x-axis label
	ylabel={load [A]},            % y-axis label
	legend style={font=\small},		
	legend pos=north west,         % legend position on plot
	legend cell align=left,        % text alignment within legend
	%domain=20:180,                 % domain for plotted functions (not needed for scatter data)
	%samples=200,                   % plot 200 samples
	]
	\addplot+ [mark=none] table [col sep=comma,row sep=newline] {Simulation_I1.dat}; % add the first plot
	\addlegendentry{$d_1$}; % add the first plot's legend entry
	\addplot+ [mark=none] table [col sep=comma,row sep=newline] {Simulation_I2.dat};
	\addlegendentry{$d_2$};
	\addplot+ [mark=none] table [col sep=comma,row sep=newline] {Simulation_I3.dat};
	\addlegendentry{$d_3$};
	%\addlegendentry{Sammon Mapping};
	%\addplot[mark=triangle,green] {x^2/200 + rand*x/1.5};
	%\addlegendentry{Squared Stress};
	\end{axis}
	\draw [decorate,decoration={brace,amplitude=4pt,mirror,raise=4pt},yshift=0pt]
	(1.24,1.6) -- (1.24,2.37) node [black,midway,xshift=0.65cm] {\small$\Delta d_1$};
	\end{tikzpicture}%
	~%
	\begin{tikzpicture}
	\pgfplotsset{%
		width=0.25\linewidth,
		height=0.25\linewidth,
		legend image code/.code={
			\draw[mark repeat=2,mark phase=2]
			plot coordinates {
				(0cm,0cm)
				(0.15cm,0cm)        %% default is (0.3cm,0cm)
				(0.3cm,0cm)         %% default is (0.6cm,0cm)
			};%
		}
	}
	\begin{axis}[
	%only marks,                    % no lines
	every axis plot/.append style={thick},
	xmin=0.3, xmax=1,              % x-axis limits
	ymin=-0.5, ymax=1.5,              % y-axis limits
	ylabel style={at={(axis description cs:0.05,0.5)}},
	xlabel={time [s]},      % x-axis label
	ylabel={flux [A]},            % y-axis label
	legend style={font=\small},
	legend pos=north west,         % legend position on plot
	legend cell align=left,        % text alignment within legend
	%domain=20:180,                 % domain for plotted functions (not needed for scatter data)
	%samples=200,                   % plot 200 samples
	]
	\addplot+ [mark=none,green!70!black] table [col sep=comma,row sep=newline] {Simulation_F12.dat}; % add the first plot
	\addlegendentry{$\xi_{21}$}; % add the first plot's legend entry
	\addplot+ [mark=none,purple] table [col sep=comma,row sep=newline] {Simulation_F23.dat};
	\addlegendentry{$\xi_{32}$};
	%\addlegendentry{Sammon Mapping};
	%\addplot[mark=triangle,green] {x^2/200 + rand*x/1.5};
	%\addlegendentry{Squared Stress};
	\end{axis}
	\end{tikzpicture}
	\caption{Voltage, input, load and flux trends of cluster $\mathcal{C}_1$ during Algorithm 1 simulation on the microgrid model $\mathbf{G}$.}
	\label{Simulation}
\end{figure*}

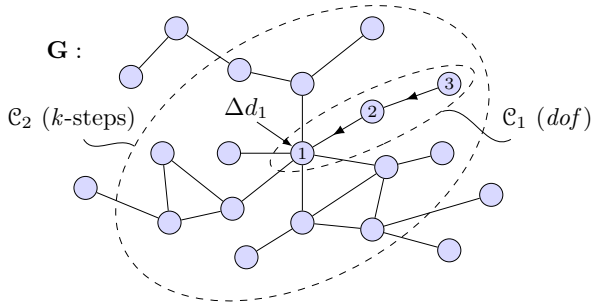
\begin{figure}
	\centering
	\scalebox{0.92}{%
		\begin{tikzpicture}[auto,
	node_style/.style={circle,draw,fill=blue!15!},
	edge_style/.style={draw=black},]
	
	\tikzset{middlearrow/.style={
			decoration={markings,
				mark= at position 0.6 with {\arrow[scale=1.2]{#1},} ,
			},
			postaction={decorate}
		}
	}
	
	\node[node_style] (v1) at (0,0) {};
    \node[node_style] (v2) at (0,-1) {};
    \node[node_style] (v3) at (1.2,-0.2) {};
    \node[node_style] (v4) at (2,0) {};
    \node[node_style] (v5) at (1,-1.1) {};
    \node[node_style] (v6) at (2.1,-1.4) {};
    \node[node_style] (v7) at (-0.8,-1.5) {};
    \node[node_style] (v8) at (1,0.6) {};
    \node[node_style] (v9) at (2.1,1) {};
    \node[node_style] (v10) at (0,1) {};
    \node[node_style] (v11) at (-1.05,0) {};
    \node[node_style] (v12) at (-1,-0.8) {};
    \node[node_style] (v13) at (-1.9,-1) {};
    \node[node_style] (v14) at (1,1.8) {};
    \node[node_style] (v15) at (-0.9,1.2) {};
    \node[node_style] (v16) at (-2,0) {};
    \node[node_style] (v17) at (-1.8,1.8) {};
    \node[node_style] (v18) at (-2.45,1.1) {};
    \node[node_style] (v19) at (-3.1,-0.5) {};
    \node[node_style] (v20) at (2.7,-0.6) {};
    
	\node at (v1) {\scriptsize$1$};
	%\node at (v2) {\scriptsize$2$};
	%\node at (v3) {\scriptsize$3$};
	%\node at (v4) {\scriptsize$4$};
	%\node at (v5) {\scriptsize$5$};
	%\node at (v6) {\scriptsize$6$};
	%\node at (v7) {\scriptsize$7$};
	\node at (v8) {\scriptsize$2$};
	\node at (v9) {\scriptsize$3$};
	%\node at (v10) {\scriptsize$10$};
	%\node at (v11) {\scriptsize$11$};
	%\node at (v12) {\scriptsize$12$};
	%\node at (v13) {\scriptsize$13$};
	%\node at (v14) {\scriptsize$14$};
	%\node at (v15) {\scriptsize$15$};
	%\node at (v16) {\scriptsize$16$};
	%\node at (v17) {\scriptsize$17$};
	%\node at (v18) {\scriptsize$18$};
	%\node at (v19) {\scriptsize$19$};
	%\node at (v20) {\scriptsize$20$};

	\draw[edge_style]  (v1) edge node{} (v2);
	\draw[edge_style]  (v1) edge node{} (v3);
	\draw[edge_style]  (v2) edge node{} (v3);
	\draw[edge_style]  (v3) edge node{} (v5);
	\draw[edge_style]  (v2) edge node{} (v5);
	\draw[edge_style]  (v5) edge node{} (v6);
	\draw[edge_style]  (v3) edge node{} (v4);
	\draw[edge_style]  (v2) edge node{} (v7);
	\draw[middlearrow={latex}]  (v8) -- node{} (v1);
	\draw[middlearrow={latex}]  (v9) -- node{} (v8);
	\draw[edge_style]  (v1) edge node{} (v10);
	\draw[edge_style]  (v1) edge node{} (v11);
	\draw[edge_style]  (v1) edge node{} (v12);
	\draw[edge_style]  (v12) edge node{} (v13);
	\draw[edge_style]  (v10) edge node{} (v15);
	\draw[edge_style]  (v15) edge node{} (v17);
	\draw[edge_style]  (v10) edge node{} (v14);
	\draw[edge_style]  (v17) edge node{} (v18);
	\draw[edge_style]  (v13) edge node{} (v16);
	\draw[edge_style]  (v12) edge node{} (v16);
	\draw[edge_style]  (v13) edge node{} (v19);
	\draw[edge_style]  (v5) edge node{} (v20);

	\draw[dashed,rotate around={-60:(v1)}] (v1) ellipse (1.8cm and 2.9cm);
	\draw[dashed,rotate around={-65:(1,0.5)}] (1,0.5) ellipse (0.4cm and 1.6cm);
	
	\node (c1) at (3.5,0.5) {$\mathcal{C}_1$ (\textit{dof})};
	\node (c2) at (-3.3,0.5) {$\mathcal{C}_2$ (\textit{k}-steps)};
	\node at (-3.4,1.5) {$\mathbf{G}:$};
	\node (d) at (-0.8,0.65) {$\Delta d_1$};
	\draw [-{latex}] (-0.65,0.45) to (v1);
	
	\draw [very thin] (2,0.6) to[in=210,out=20] (c1);
	\draw [very thin] (-2.4,0.1) to[in=300,out=145] (c2);
	
	%\node at (-2.5,1) {$\mathbf{G}:$};
	%\node at (-1.5,2.4) {$\mathcal{C}_1$};
	%\node at (1.4,2.4) {$\mathcal{C}_2$};
	\end{tikzpicture}}
\caption{The \textit{dof}-based algorithm and the $k$-steps reachability set algorithm are applied to mG model $\mathbf{G}$, selecting clusters $\mathcal{C}_1$ and $\mathcal{C}_2$ to contain disturbance $\Delta d_1$.} 
\label{Cluster simulation}
\end{figure}

Here we show an application of Algorithm \ref{Algorithm} to the problem of voltage regulation in DC microgrids. The experiment is conducted with the MATLAB Power Systems Toolbox. We rely on a converter-based mG model (\cite{Cucuzzella,Michele,AndreaSecondaryControl}), represented by a set of $n$ nodes that can exchange currents through a graph $\mathbf{G}$ (see Fig. \ref{Cluster simulation}). Each node hosts a dynamical system of the form

\begin{align*}
	\dot{x}_i & = f_i\left( x_i,u_i,\xi_i,d_i \right) \\
	        & = \begin{bmatrix}
	        0 & 1/C_i \\ -1/L_i & -R_i/L_i
	        \end{bmatrix} x_i + \begin{bmatrix}
	        0 \\ V_{in,i}
	        \end{bmatrix} u_i + \begin{bmatrix}
	        1 \\ 0
	        \end{bmatrix} \xi_i - \begin{bmatrix}
	        1 \\ 0
	        \end{bmatrix} d_i,
\end{align*}

where $x_i = [ V_i \;\: I_i ]'$ is the vector state comprising the converter output voltage and internal current, $u_i$ is the input (duty cycle), $d_i$ is the load disturbance. The coupling current $\xi_i$ is defined as in \eqref{Coupling} with $\xi_{ij} = G_{ij}(V_j-V_i)$, where $G_{ij}$ is the conductance of the corresponding transmission line. The other parameters $C_i$, $L_i$, $R_i$, $V_{in,i}$ are the capacitance, inductance, resistance and input voltage of the converter, respectively. Each node is equipped with a state-feedback map $u_i = g_i(x_i,x_i^{r},\xi_i,d_i)$ that, in nominal conditions, guarantees perfect tracking at steady state, $x_i = x_i^{r}$. The map $g_i$ can be synthesized according to different methods that can be found, for instance, in the previously mentioned literature. For the following simulation, we use the microgrid electrical parameters in Table I of \cite{AndreaSecondaryControl} and we implement the linear control map described therein. 

We consider the network to be in its steady-state condition, until a disturbance variation $\Delta d_1$ affects node 1 at time 0.6s. Algorithm 1 explores the node space by selecting the nodes that increase cluster \textit{dof} and maximize an availability measure assigned to each node, defined as 

\begin{equation*}
	\Psi_{i} = |d_i|(1-|u_i-0.5|),
\end{equation*} 

that quantifies the control balance (deviation from $50\%$ duty cycle) weighted over the disturbance magnitude. After 2 iterations, the $dof$-based algorithm and the $k$-steps reachability set algorithm select clusters $\mathcal{C}_1$ and $\mathcal{C}_2$ respectively (Fig. \ref{Cluster simulation}) and solve the local optimization problem \eqref{optimization problem}. We note that the $dof$-based exploration method is able to contain the disturbance in a much smaller community. Fig. \ref{Simulation} displays the output voltage, input, load and exchanged flux among the three nodes in cluster $\mathcal{C}_1$. After the disturbance variation $\Delta d_1$, nodes 2 and 3 increase their output voltage to generate a local flux redistribution that benefits node 1. The output voltage of node 1, on the other hand, quickly returns to a pre-disturbance value, such that nodes external to $\mathcal{C}_1$ do not perceive any flux variation.

%%%%%%%%%%%%%%%%%%%%%%%%%%%%%%%%%%%%%%%%%%%%%%%%%%%%%%%%%%%%%%%%%%%%%%%%%%%%%%

\section{CONCLUSIONS}\label{Section Conclusion}

In this paper, we tackled the problem of local disturbance containment in networked dynamical systems. We introduced a novel clustering measure, the \textit{dof}, that expresses the structural availability of a cluster to contain a disturbance. To reduce the computational effort to evaluate a cluster according to our measure, we proved that the diagonal blocks of a Laplacian matrix corresponding to connected clusters are nonsingular. Finally, we defined a greedy clustering algorithm and showed its applicability in the context of microgrids voltage control.

Many interesting aspects of this clustering approach still need to be explored. As a first step, the \textit{dof} measure could be employed to generate global partitions of the network. The exact formulation of this problem is computationally prohibitive in principle, since the number of possible partitions of a set scales according to the Bell numbers. Secondly, our \textit{dof} measure could be strengthened by adding an additional structural condition on flux circulation: each node in the cluster must be reachable by the flux redistribution enforced by the modification of the references. Another promising extension can be achieved by solving the optimization problem \eqref{optimization problem} within the framework of game theory, letting the nodes negotiate their own references as a value assignment problem. Finally, thanks to the general formulation of the problem, it seems valuable to apply the \textit{dof} concept to different domains, such as hydro power plants or traffic networks.

\section*{ACKNOWLEDGEMENTS}
A great thanks to A. La Bella and R. Scattolini for the stimulating discussions on microgrids clustering.

%\bibliographystyle{ifacconf}

           % bib file to produce the bibliography
                                              % with bibtex (preferred)
                                                  
\end{document}